# Capacity Enhancement Analysis and Implementation of a 3D Array Based on Miniaturized Dipole Antennas

Yongzheng Li, *Student Member, IEEE,* and Wanchen Yang, *Senior Member, IEEE,* Shuai S. A. Yuan, *Student Member, IEEE,* Zhitao Ye, *Student Member,* Chongwen Huang, *Member, IEEE,* Xiaoming Chen, *Senior Member, IEEE,* Wenquan Che, *Fellow, IEEE,* and Wei E. I. Sha, *Senior Member, IEEE*

*Abstract*—Theoretically, the three-dimensional (3D) array architecture provides a higher communication degree of freedom (DoF) compared to the planar arrays, allowing for greater capacity potential in multiple-input multiple-output (MIMO) systems. However, in practical implementations, the upper elements of 3D arrays significantly degrade the performance of the lower elements, leading to increased inter-element correlation and reduced array efficiency. As a result, the expected enhancement in MIMO performance is often suboptimal. To address this issue, this work employs a miniaturized antenna element to reduce the inter-element correlation and thus enhance the DoF of the 3D array. Moreover, to mitigate the efficiency degradation of the lower elements caused by the upper ones, the structures of lower elements are modified to achieve wideband impedance matching. The influence of upper element profile distribution on DoF and element efficiency is investigated, and the scalability of the proposed 3D array is theoretically analyzed. Finally, the MIMO performance of the proposed 3D array is evaluated under 3GPP scenarios, demonstrating a 16% higher capacity than conventional 2D arrays under the same SNR of 20 dB and a physical aperture area of 6.26 $\lambda_0^2$. These results indicate that 3D arrays of appropriately arranged miniaturized elements offer a promising approach to enhancing MIMO system performance.

This work was supported in part by the National Natural Science Foundation of China under Grant 62471219, and 62321002. (*Wanchen Yang, W. E. I. Sha and Wenquan Che contributed equally to this work.*) (*Corresponding authors: Wenquan Che*).

Y. Li and W. Che are with the Guangdong Provincial Key Laboratory of Millimeter-Wave and Terahertz, Guangdong-Hong Kong-Macao Joint Laboratory for Millimeter-Wave and Terahertz, School of Electronic and Information Engineering, South China University of Technology, Guangzhou 510641, China. (e-mail: eelyz@ieee.org, eewqche@scut.edu.cn).

W. Yang and Z. Ye are with the College of Electronic and Information Engineering, Nanjing University of Aeronautics and Astronautics, Nanjing 211106, China. (e-mail: wcyang@nuaa.edu.cn).

S. S. A. Yuan is with the Department of Electronics and Nano engineering, School of Electrical Engineering, Aalto University, 02150 Espoo, Finland.

W. E. I. Sha and C. Huang are with the College of Information Science and Electronic Engineering, Zhejiang University, Hangzhou 310027, China (e-mail: weisha@zju.edu.cn).

X. Chen is with the School of Information and Communication Engineering, Xi'an Jiaotong University, Xi'an 710049, China.

## I. INTRODUCTION

MIMO systems utilize spatial multiplexing to enhance spectral efficiency [1], making them a fundamental technology for high-capacity access in the B5G and even 6G era. Since the base station sites are limited, it is desirable to maximize the communication capacity of a single MIMO array to reduce deployment costs. To this end, more antenna elements must be integrated within a constrained physical aperture to increase the number of available channels. However, this densification inevitably reduces the inter-element spacing.

Since the communication capacity of a MIMO array depends on both beamforming gain and DoF gain [2], reducing antenna spacing has two main effects. First, it increases mutual coupling between antenna elements, thereby reducing array efficiency [3] and consequently lowering beamforming gain. Second, it raises correlation among elements, which diminishes the additional DoF gain as the number of elements grows [4].

Several studies have proposed decoupling structures to mitigate mutual coupling and enhance antenna efficiency [5-8]. However, these structures are inherently frequency-dependent, leading to a limited bandwidth. Furthermore, in practical MIMO base station applications, the angular power spectrum of incoming waves is confined within a limited angular spread [9]. As a result, improving isolation does not necessarily reduce correlation [10]. Therefore, decoupling structures are not an effective solution for improving the communication capacity of MIMO base station arrays.

Another approach to enhance the DoF involves loading scattering structures above the antenna array [10-12]. A phase correction element (PCE) was introduced in [10] to adjust the near-field phase distribution, thereby reducing inter-element correlation. However, the required aperture phase shift distribution varies among different elements, making it difficult to find an optimal configuration that minimizes correlation for all elements, thus limiting its scalability to large-scale arrays. A high-permittivity dielectric material was loaded above the antenna array in [12] to induce an equivalent phase center shift, thereby improving DoF. However, the enhancement in DoF achieved by this method is rather limited. Additionally, the magnitude of the phase center shift is positively correlated with the dielectric constant, and the use of high-permittivity materials may generate surface waves, ultimately degrading the beamforming gain.

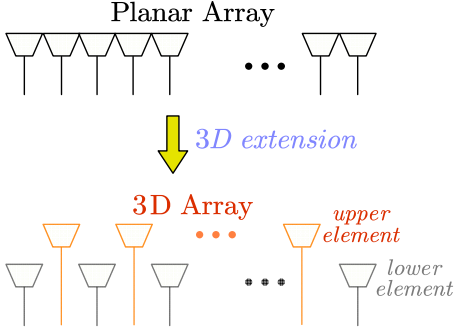

Fig. 1. Sketch of the 3D array architecture.

Recently, an alternative approach has been proposed to enhance the communication capacity of MIMO arrays by utilizing a three-dimensional (3D) array architecture [13]. As shown in Fig. 1, in a 3D array, certain antenna elements are elevated, increasing the physical spacing between adjacent elements and thereby reducing mutual coupling. Meanwhile, the DoF is fundamentally limited by the surface area enclosing the antenna array [4], [14], which is also consistent with the equivalence principle in electromagnetic theory. Consequently, the 3D array provides a larger effective surface area than conventional planar arrays, allowing for a higher DoF limit. However, the elevated elements introduce shielding effects on the lower ones, resulting in distorted radiation patterns and insignificant DoF improvement compared to planar arrays. Furthermore, lifting the upper elements alters the radiation environment, potentially causing impedance mismatch and reducing array efficiency. As a result, these effects limit the achievable communication capacity improvement in 3D arrays.

To address these issues, this work proposes a 3D array that employs wideband miniaturized antenna elements. By miniaturizing the antenna element, radiation pattern distortion in the lower elements is effectively mitigated, leading to an increased DoF. Additionally, the lower elements are structurally optimized to improve impedance matching, which significantly enhances the overall array efficiency. Besides, the influence of upper element profile distribution on DoF and element efficiency is investigated, and the scalability of the proposed 3D array is theoretically analyzed. It is shown that as the array scale increases, the average element efficiency and DoF gain per-element asymptotically approach limiting values of approximately 70% and 0.35, respectively. Overall, the proposed 3D array overcomes the fundamental limitations on DoF and efficiency imposed by aperture constraints, and achieves notable capacity improvements under a 3GPP channel model compared to its planar counterpart.

The remainder of this manuscript is organized as follows. Section II clarifies the issues of efficiency degradation and DoF deterioration in 3D arrays using the conventional elements. Section III presents a 3D array with miniaturized elements to address these issues. Section IV examines the impact of various upper element profile distributions on both DoF and efficiency, and then analyzes the scalability of the proposed 3D array. Section V evaluates the MIMO performance under 3GPP channel scenarios. Finally, Section VI concludes the work.

## II. THEORETICAL ANALYSIS AND PROBLEM STATEMENT

In this section, the DoF of a conventional dual-polarized crossed-dipole 3D array is analyzed using the *Kronecker* model. The results show that conventional dipole elements suffer from low efficiency and radiation pattern distortion in the 3D array architecture, leading to limited DoF improvement compared to planar arrays.

### A. Theoretical Analysis

The DoF describes the number of independent antenna ports in an array and serves as an intuitive metric for evaluating spatial-multiplexing performance of an antenna array in MIMO system. To compute the DoF, the antenna array is analyzed using the *Kronecker* model. The covariance matrix $\mathbf{R}$ taking into account the impact of antenna efficiency is the *Hadamard* product of the correlation matrix $\mathbf{\Phi}$ and the efficiency matrix $\mathbf{\Xi}$ [13], [15]–[17].

$$\mathbf{R} = \mathbf{\Phi} \cdot \mathbf{\Xi} \tag{1}$$

The correlation matrix $\mathbf{\Phi}$ can be obtained by

$$\mathbf{\Phi} = \begin{pmatrix} 1 & \rho_{1,2} & \cdots & \rho_{1,N_r} \\ \rho_{1,2}^* & 1 & \cdots & \rho_{2,N_r} \\ \vdots & \vdots & \ddots & \vdots \\ \rho_{1,N_r}^* & \rho_{2,N_r}^* & \cdots & 1 \end{pmatrix} \tag{2}$$

where

$$\rho_{mn} = \frac{\oint G_{mn}(\Omega)d\Omega}{\sqrt{\oint G_{mm}(\Omega)d\Omega \cdot \oint G_{nn}(\Omega)d\Omega}} \tag{3}$$

with

$$G_{mn}(\Omega) = \kappa E_{\theta m}(\Omega)E_{\theta n}^*(\Omega)P_\theta(\Omega) + E_{\phi m}(\Omega)E_{\phi n}^*(\Omega)P_\phi(\Omega) \tag{4}$$

$E_{\theta m}(\Omega)$ and $E_{\phi m}(\Omega)$ represent the $\theta$ and $\varphi$ E-field components, respectively. $\kappa$ is the crossed polarization discrimination, which is taken as 1 to simulate the polarization balanced situation. $P(\Omega)$ is the angular power spectrum, which characterizes the distribution of the incident wave energy as a function of the spatial angle $\Omega$. Notably, $P(\Omega)$ is assumed uniformly distributed within the angular spread $|\theta| \leq \theta_0$ [18]. Meanwhile, the efficiency matrix $\mathbf{\Xi}$ can be calculated using

$$\mathbf{\Xi} = \sqrt{\mathbf{e}}\sqrt{\mathbf{e}}^T \tag{5}$$

where

$$\mathbf{e} = \begin{pmatrix} e_1^{\text{emb}} & e_2^{\text{emb}} & \cdots & e_{N_r}^{\text{emb}} \end{pmatrix}^T \tag{6}$$

with

$$e_n^{\text{emb}} = 1 - |S_{1n}|^2 - |S_{2n}|^2 - \cdots - |S_{N_r,n}|^2 \tag{7}$$

After obtaining the covariance matrix **R** of the antenna array, the DoF is computed using [19]

$$DoF = \left(\frac{tr(\mathbf{R})}{\|\mathbf{R}\|_F}\right)^2 \qquad (8)$$

where $tr(\cdot)$ and $\|\cdot\|_F$ denotes the trace and *Frobenius* norm of the matrix, respectively.

*B. Problem Statement*

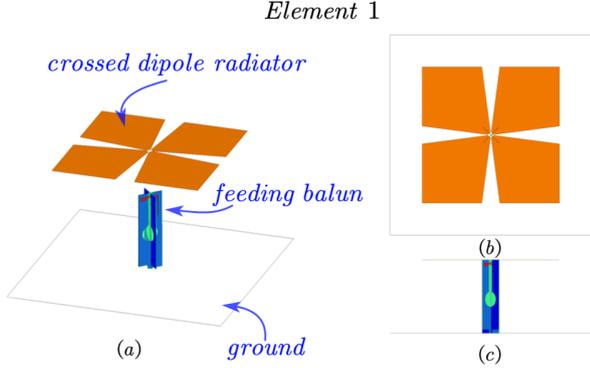

Fig. 2. Configuration of the *Element* 1: (a) 3D view, (b) Top view, (c) Side view.

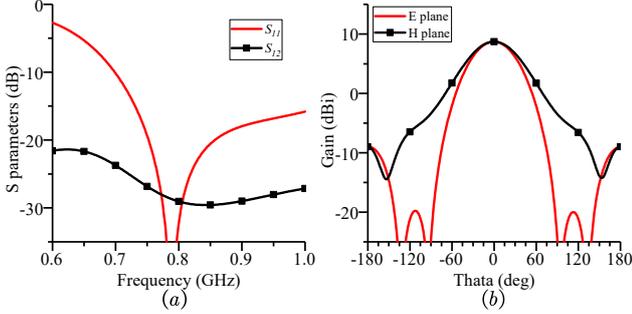

Fig. 3. (a) *S* Parameter and (b) Radiation pattern of the *Element* 1.

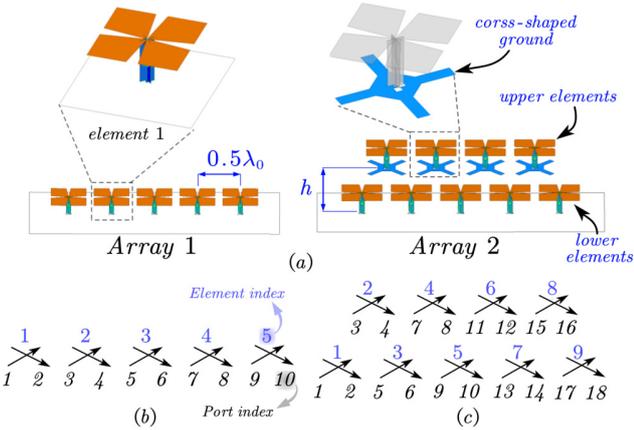

Fig. 4. (a) Configuration of *Array* 1 and *Array* 2. Element and port index for (b) *Array* 1 and (c) *Array* 2.

To highlight the challenges in 3D array design, a 3D array constructed using conventional crossed dipoles is first analyzed. The configuration of the conventional crossed dipole, refer to as *Element* 1, is illustrated in Fig. 2. Fig. 3(a) shows the *S* parameters of the *Element* 1. As observed, the reflection coefficient is below −10 dB within the frequency range of 0.7 GHz to 1 GHz, while the isolation between the two orthogonal polarizations exceeds 23 dB. Fig. 3(b) illustrates the radiation pattern of the *Element* 1, where a maximum gain of 8.7 dBi is observed in the broadside. As shown in Fig. 4, a planar array and a 3D array are constructed using *Element* 1, referred to as *Array* 1 and *Array* 2, respectively. Specifically, the spacing in *Array* 1 is set to 0.5 $\lambda_0$, where $\lambda_0$ is the free-space wavelength at 825 MHz. *Array* 2 is formed by adding four upper elements to *Array* 1, with the upper elements placed at a profile of $h$=0.5 $\lambda_0$. In addition, a cross-shaped ground is placed beneath the upper elements to ensure their proper operation.

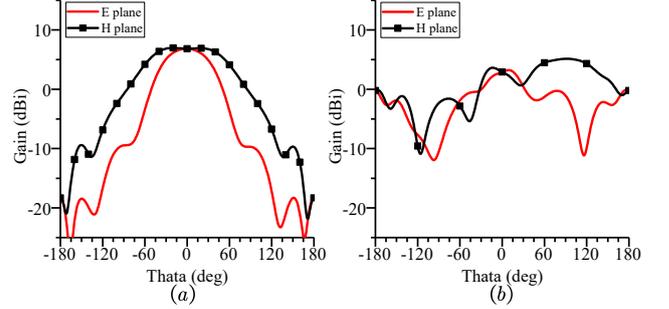

Fig. 5. Radiation patterns of: (a) *Port* 5 in *Array* 1 and (b) *Port* 9 in *Array* 2.

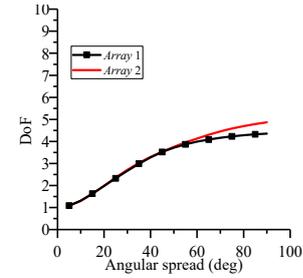

Fig. 6. DoF of *Array* 1 and *Array* 2 at 825 MHz.

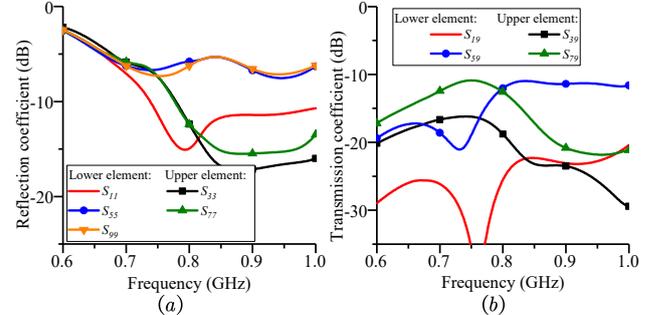

Fig. 7. *S* parameters of the *Array* 2: (a) Reflection coefficient, (b) Transmission coefficient of *Port* 9.

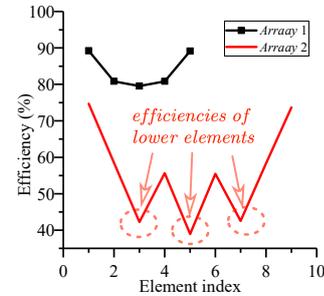

Fig. 8. Elements efficiency of *Array* 1 and *Array* 2 at 825 MHz.

Fig. 5 compares the radiation patterns of the central port — *Port* 5 in *Array* 1 and *Port* 9 in *Array* 2, respectively. As observed, the lower elements in *Array* 2 suffers significant radiation pattern distortion compared to those in *Array* 1, with the peak gain degrading from 8.7 dBi to 2.9 dBi. This degradation results in increased mutual correlation between antenna elements. The DoF for the *Array* 1 and *Array* 2 is shown in Fig. 6. As observed, the DoF improvement achieved by *Array* 2 over *Array* 1 is marginal. Even under an angular spread of θ=90°, the DoF improvement is only 11.68%. These results indicate that the shielding effect introduced by the upper elements restricts the DoF enhancement in the 3D array architecture.

Meanwhile, Fig. 7 illustrates the *S* parameters of the *Array* 2. As observed, the reflection coefficient of the lower elements in *Array* 2 deteriorates to above -6 dB, while the isolation between elements decreases to approximately 10 dB. These results suggest that the presence of upper elements alters the radiating environment of the lower elements, leading to impedance mismatch. Additionally, part of the energy radiated by the lower elements is reflected by the upper elements, further reducing the isolation between adjacent lower elements. The deterioration in the *S* parameters directly impacts array efficiency. As shown in Fig. 8, at 825 MHz, the efficiency of *Array* 1 remains above 80%, whereas that of the lower elements in *Array* 2 drops to as low as 39%.

Although the 3D array architecture theoretically provides a higher DoF than planar arrays, practical implementation faces challenges due to shielding and scattering effects introduced by the upper elements, which degrade both array efficiency and DoF gain. Consequently, the improvement in communication capacity achieved by deploying a 3D array in MIMO systems is significantly constrained. Therefore, to fully realize the potential of the 3D array architectures, it is essential to minimize the interference between the upper and lower elements.

## III. DoF and Efficiency Enhancement of the 3D Array Based on Miniaturized Elements

In this section, a miniaturized antenna is employed to mitigate the DoF degradation in 3D arrays. The resonances of the lower elements are subsequently adjusted to enhance array efficiency. As a result, a 3D array with high DoF—exceeding the physical limits imposed by the aperture of planar arrays—and an efficiency above 73% is achieved.

### A. Enhancement of the DoF with the Miniaturized Elements

Miniaturized antennas are characterized by a small aperture and reduced shielding effects. To mitigate the influence of upper-layer elements on the lower ones, a miniaturized antenna element proposed in [20], referred to as *Element* 2, is adopted in the 3D array, as shown in Fig. 9(a). The configuration of the resulting miniaturized-element 3D array, denoted as *Array* 3, is illustrated in Fig. 9(b). Notably, *Array* 3 shares the same architecture as *Array* 2, with *Element* 1 replaced by *Element* 2.

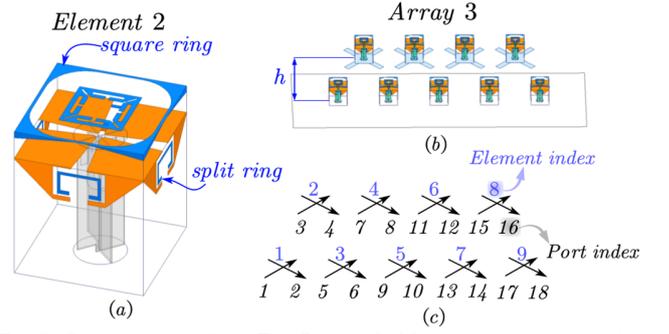

Fig. 9. Configurations of: (a) The *Element* 2 (Miniaturized antenna) and (b) The *Array* 3. (c) Element and port index of *Array* 3.

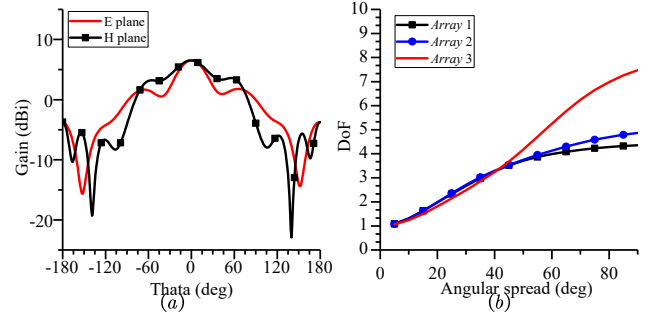

Fig. 10. (a) Radiation pattern of the central *Port* 9 in *Array* 3. (b) DoF for *Array* 1, *Array* 2 and *Array* 3 at 825 MHz.

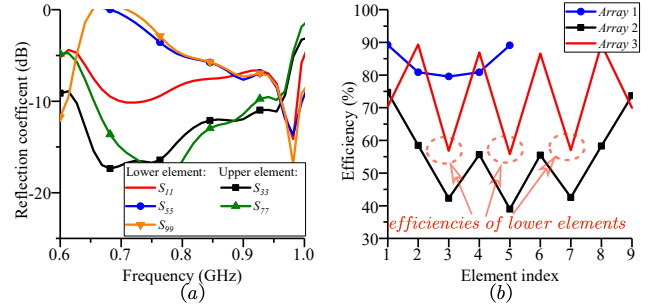

Fig. 11. (a) Reflection coefficients of *Array* 3, (b) Efficiency for each element of the *Array* 1, *Array* 2 and *Array* 3 at 825 MHz.

The radiation pattern of the central *Port* 9 in *Array* 3 is shown in Fig. 10(a). As observed, the beam direction is restored to broadside with a peak gain of 6.5 dBi, indicating that the use of miniaturized elements enhances the DoF in the 3D array architecture. To further validate this, Fig. 10(b) compares the DoF of *Array* 1, *Array* 2 and *Array* 3 at 825 MHz. The results show that when the angular spread exceeds 45°, *Array* 3 achieves a notable DoF enhancement over *Array* 2. Especially, at θ=90°, the DoF of *Array* 3 is 55.6% higher than that of *Array* 2. Moreover, the half-space DoF limit of a planar array with the same physical aperture size as *Array* 3 is 6.14 at 825 MHz, while the DoF of *Array* 3 reaches 6.26 at θ=70°, confirming the feasibility of using a 3D array architecture to surpass the DoF limit of planar arrays.

However, the presence of upper elements alters the electromagnetic environment of the lower elements, leading to impedance mismatch. As shown in Fig. 11(a), the reflection coefficients of the lower elements remain suboptimal, with *Port* 9 in particular exhibiting a relatively poor value of approximately -5.5 dB at 825 MHz. This mismatch contributes to reduced array efficiency. As illustrated in Fig. 11(b), the efficiencies of the lower elements (elements 3, 5, and 7) drop below 60%, which in turn degrades the beamforming gain.

### B. Enhancement of the Array Efficiency by Resonant Modes Adjustments

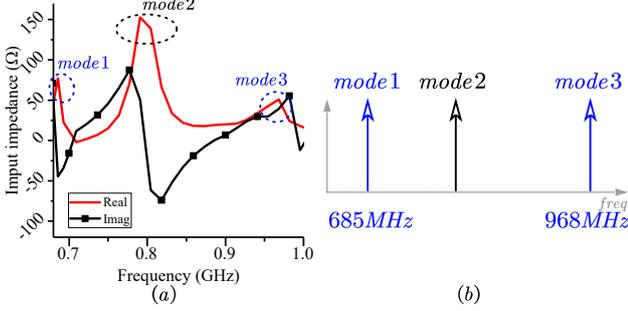

Fig. 12. (a) Input impedance and (b) Resonant mode distribution of the *Array* 3.

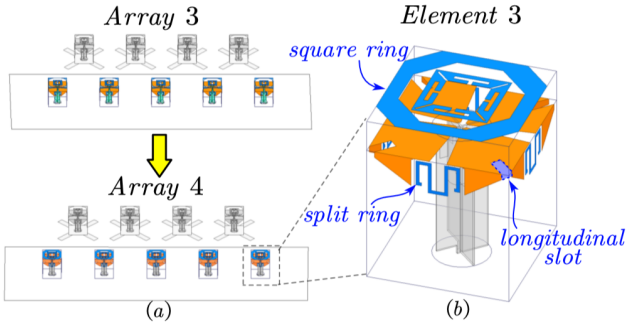

Fig. 13. Configurations of (a) *Array* 4 and (b) The *Element* 3.

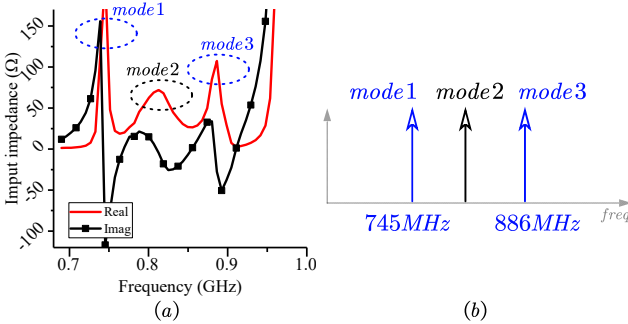

Fig. 14. (a) Input impedance and (b) resonant mode distribution of the *Array* 4.

As previously discussed, the impedance mismatch significantly reduces the efficiency of the lower elements. To address this issue, it is essential to consider the influence of the upper elements in the design of the lower elements. To this end, a detailed analysis of the resonance mode distribution of the lower elements was conducted, as shown in Fig. 12. It is observed that the lower elements in *Array* 3 exhibit three distinct resonance modes. However, the large frequency spacing between these modes results in substantial impedance fluctuations, thereby causing mismatch. To reduce the mode spacing, two structural modifications are implemented based on the approach proposed in [20]. First, the dimension of the square ring is reduced and chamfered corners are introduced to shift the *mode* 1 to a higher frequency. Second, the *mode* 3 is shifted to a lower frequency by extending the current path of the split rings and incorporating longitudinal slots into the dipole arms. The resulting structure of the optimized element is illustrated in Fig. 13 and is referred to as *Element* 3. The input impedance of the *Element* 3 is shown in Fig. 14, where it is observed that *modes* 1 and 3 are brought closer to mode 2, leading to a more uniform impedance profile.

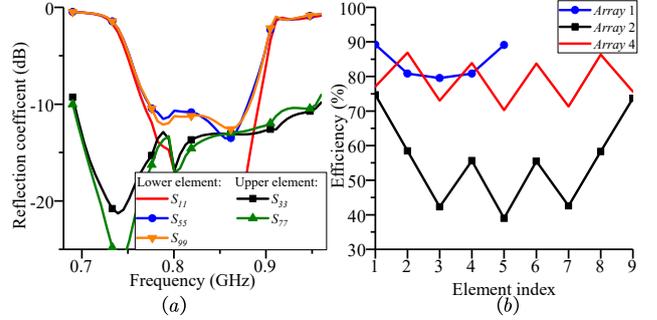

Fig. 15. (a) Reflection coefficients of *Array* 4 and (b) Element efficiency at 825 MHz of the *Array* 1,2 and 4.

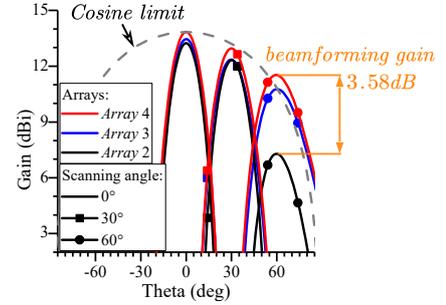

Fig. 16. Scanning beams at 0°, 30° and 60° for *Array* 2, *Array* 3 and *Array* 4 at 825 MHz.

Subsequently, *Array* 3 is further improved by replacing its lower elements with *Element* 3, resulting in *Array* 4. As shown in Fig. 15(a), the reflection coefficient of the lower elements in *Array* 4 remains below –10 dB over the 770-880 MHz range. The corresponding element efficiencies, illustrated in Fig. 15(b), exceed 73% for all elements, reaching levels comparable to those of *Array* 1. To clearly demonstrate the improvement in beamforming gain, Fig. 16 shows the scanning beams at a 60° angle for *Array* 2, *Array* 3, and *Array* 4. As observed, *Array* 4 exhibits a notable beamforming gain improvement of more than 3 dB over *Array* 2. Moreover, the theoretical gain limit of a planar array is given by $4\pi A_e Cos\theta/\lambda^2$ [3], where $A_e$ is the effective aperture area at broadside. It is observed that the gain of *Array* 4 at a 60° scan angle exceeds this limit, confirming that the 3D array composed of miniaturized elements can surpass the fundamental aperture-constrained gain limit at wide angles. Therefore, the proposed 3D array based on miniaturized elements enables the simultaneous realization of high DoF and high beamforming gain.

### C. Fabrication and Verification

To validate the effectiveness of the *Array* 4 design, the array was fabricated and measured. The *S* parameters were measured using an R&S ZVA-40 vector network analyzer. Owing to the symmetry of the *Array* 4, only the reflection coefficients of *Ports* 1, 3, 5, 7, and 9, along with their corresponding transmission coefficients to other ports, are presented. As shown in Fig. 17 (a), the antenna array achieves a -10 dB reflection coefficient bandwidth spanning 775 MHz to 898 MHz, corresponding to a fractional bandwidth of 14.7%. Fig. 17(b–f) show the measured transmission coefficients from *Ports* 1, 3, 5, 7, and 9 to the other ports. For clarity, only the five highest transmission coefficients for each port are plotted, all of which remain below -15 dB.

The radiation pattern of *Array* 4 was measured in a 40-meter far-field chamber, as shown in Fig. 18. The measured results are given in Fig. 19, where only the $\varphi=0°$ and $\varphi=90°$ plane radiation patterns of *Ports* 1, 3, 5, 7, and 9 at 825 MHz are provided based on symmetry. The measured patterns exhibit good agreement with the simulated results.

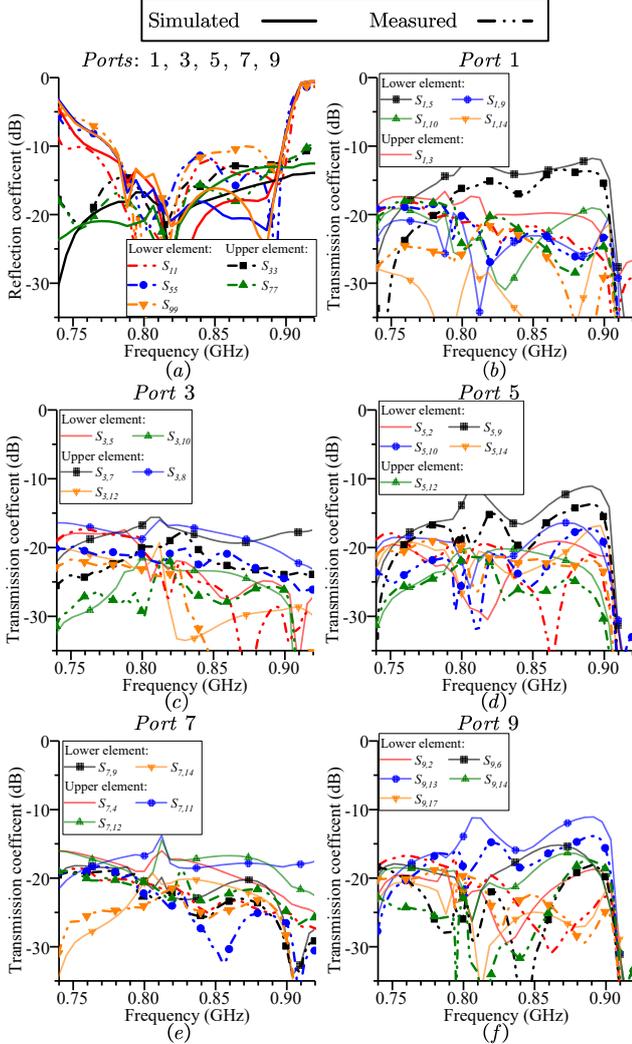

Fig. 17. Measured *S* Parameters: (a) Reflection coefficients and transmission coefficients for: (b) *Port* 1, (c), *Port* 3, (d) *Port* 5, (e) *Port* 7 and (f) *Port* 9, respectively.

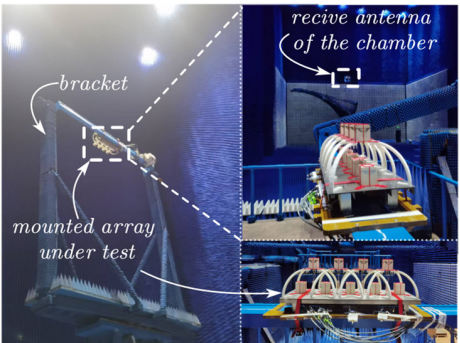

Fig. 18. Radiation pattern measurement environment.

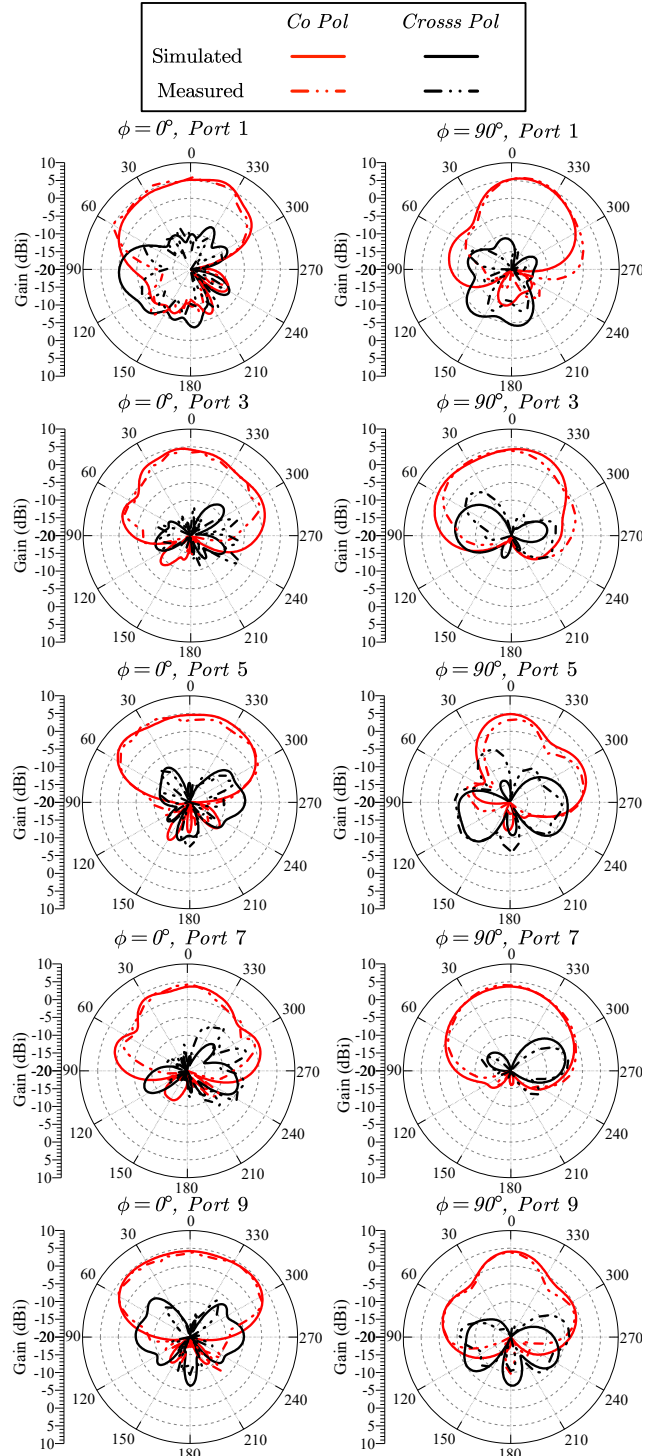

Fig. 19. Measured radiation patterns for *Ports* 1,3, 5, 7, and 9 in $\varphi=0°$ and $\varphi=90°$ plane at 825 MHz, respectively.

IV. PROFILE DISTRIBUTION AND ARRAY SCALING IN 3D ARRAYS

In the 3D array architecture, the profiles of the upper elements introduce additional design parameters. The enhancement of the DoF is influenced by the spatial distribution of these element profiles. Therefore, the impact of various profile distributions on both the DoF and array efficiency is first investigated. Furthermore, in practical massive MIMO scenarios, the array scale is significantly larger. To evaluate the scalability of the proposed design, the variations in DoF and array efficiency with respect to different array scales are analyzed.

*A. Impact of Upper Element Profile.*

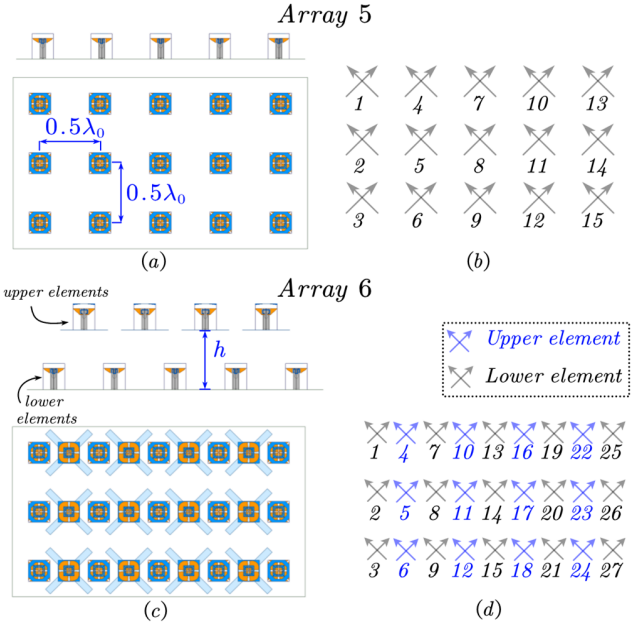

Fig. 20. (a) Configuration and (b) Element indexes of the *Array* 5. (c) Configuration and (d) Index for elements of the *Array* 6.

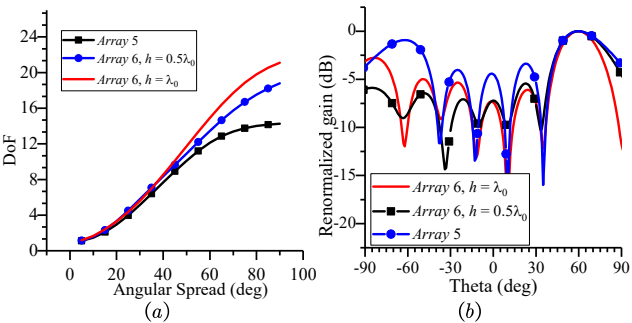

Fig. 21. Performances of *Array* 5 and *Array* 6: (a) DoF and (b) Renormalized radiation patterns when scans at 60° at 825 MHz.

In practical base station applications, the array aperture size is typically limited to 0.5 meters × 1 meter due to installation space limitations. Accordingly, a 3D array consisting of three rows with a row spacing of 0.5 $\lambda_0$ is considered for analysis, where $\lambda_0$ is the wavelength at 825 MHz. As shown in Fig. 20, a 3 × 5 planar array with an intra-row element spacing of 0.5 $\lambda_0$ is initially constructed and referred to as *Array* 5. Consequently, a 3D array is then formed by loading upper elements at a profile $h$, and this array is referred as to *Array* 6.

To investigate the impact of the profile $h$ on the DoF, Fig. 21(a) compares the DoF of *Array* 5 and *Array* 6 under two different values of $h$=0.5 $\lambda_0$ and $h$=$\lambda_0$. It is observed that increasing $h$ consistently enhances the DoF. Specifically, within an angular spread of 60°, *Array* 6 with $h$=0.5 $\lambda_0$ achieves a 10% higher DoF than *Array* 5, while with $h$=$\lambda_0$, the DoF increases by 23%. This improvement is primarily attributed to the enlarged projected aperture of the 3D array at oblique angles, which allows the formation of narrower beams and thus higher angular resolution. This effect is corroborated by Fig. 21(b), which shows that *Array* 6 with $h$=$\lambda_0$ achieves the narrowest beamwidth at a 60° scan angle.

To evaluate the impact of the upper element profile on array efficiency, Fig. 22 compares the element efficiencies of *Array* 5 and *Array* 6 for the two aforementioned profiles. The results demonstrate that array efficiency improves with increasing $h$, as greater separation between upper and lower elements helps reduce mutual coupling.

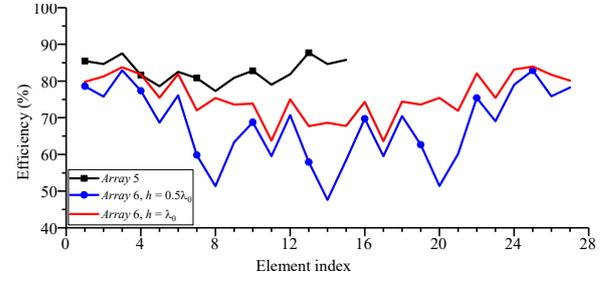

Fig. 22. Element efficiencies of *Array* 5 and *Array* 6 at 825 MHz.

*B. Impact of Non-Uniform Profile Distribution.*

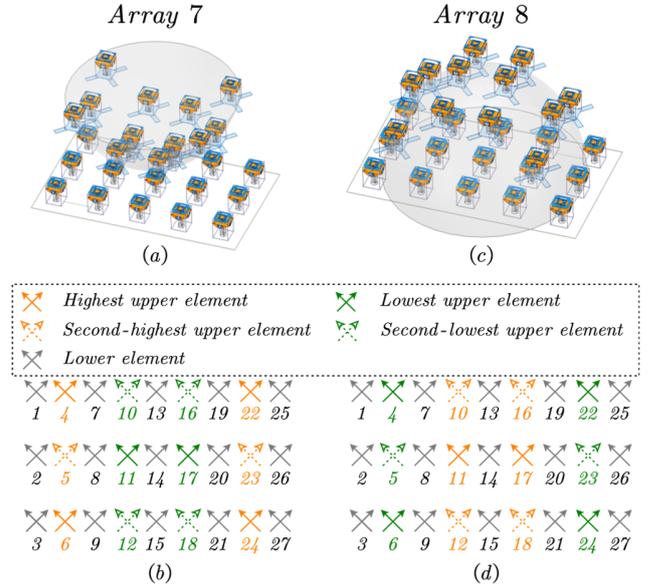

Fig. 23. (a) Configuration and (b) Elements index for *Array* 7. (c) Configuration and (b) Elements index for *Array* 8.

In the follows, the effect of non-uniform distribution of the upper element profiles is investigated. Specifically, two non-uniform upper element profile distributions are examined: one with upper elements arranged on an upper hemispherical surface and the other on a lower hemispherical surface. The corresponding configurations are illustrated in Fig. 23 and referred to as *Array* 7 and *Array* 8, respectively. To ensure a fair comparison, the average profile is maintained at $h=0.75 \lambda_0$, where $\lambda_0$ is the wavelength at 825 MHz.

Fig. 24 compares the DoF of *Array* 6, 7, and 8, all having the same average profile. As shown, *Array* 7 achieves a slightly higher DoF than *Array* 6. Meanwhile, *Array* 8 features a higher DoF than *Array* 7 within the angular range $\theta \in (0°, 60°)$, surpassing *Array* 7 by 14.5% at $\theta=35°$ in particular. However, in the higher angle range $\theta \in (60°, 90°)$, the DoF of *Array* 8 is lower than that of *Array* 7 and is nearly identical to that of *Array* 6. These results indicate that by adjusting the profile distribution, the DoF distribution can be modified in relation to the angular spread.

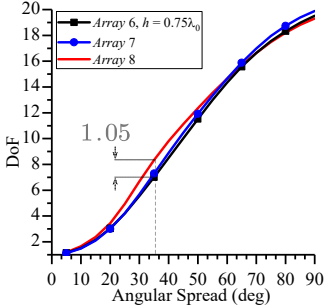

Fig. 24. DoF of the *Array* 7 and *Array* 8 at 825 MHz.

Fig. 25 examines the impact of the two non-uniform upper element profile distribution architectures on array efficiency. When the average profile is fixed at $0.75 \lambda_0$, the average efficiency difference across the arrays is minor (<3%). However, *Array* 8 shows a more gradual efficiency fluctuation across different elements compared to *Array* 6 and *Array* 7. This improvement is attributed to the mitigation of mutual coupling among the inner elements, which are typically more susceptible to coupling due to being surrounded by a greater number of neighboring elements. *Array* 8 mitigates this by elevating the profiles of surrounding upper elements, thereby reducing mutual coupling and enhancing the efficiency of the inner elements.

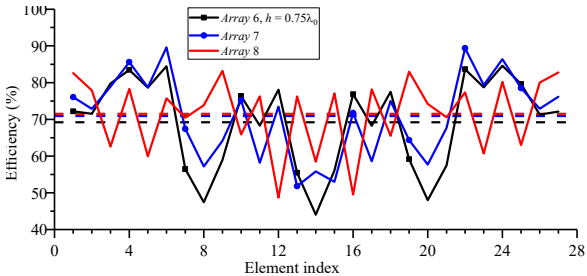

Fig. 25. Element efficiency of *Array* 6, *Array* 7 and *Array* 8 at 825 MHz.

*C. Impact of the Array Scale.*

To assess the suitability of 3D array structures for large-scale MIMO systems, it is essential to evaluate the impact of array scaling on both DoF and efficiency. To this end, a 5-row 3D array is constructed using the miniaturized elements and refer to as *Array* 9, as shown in Fig. 26. Fig. 27 compares the element efficiencies of arrays with 1, 3, and 5 rows (*Array* 4,

*Array* 6, and *Array* 9), all configured with a profile of $h=\lambda_0$ and row spacing of $0.5 \lambda_0$ to ensure a fair comparison. The results show that the average efficiency decreases with increasing array scale and eventually stabilizes. Specifically, scaling from 1 to 3 rows (*Array* 4 vs. *Array* 6) leads to an 11% reduction in the average efficiency, while further expanding to 5 rows (*Array* 6 vs. *Array* 9) results in an additional 5% decrease. Beyond this point, further scaling has minimal effect on the average efficiency. Therefore, a 5-row configuration is sufficient to represent the stabilized efficiency of large-scale 3D arrays, which converges to approximately 70%.

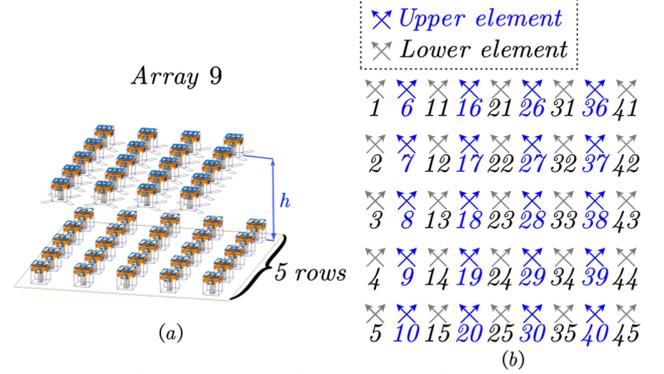

Fig. 26. (a) Configuration and (b) Elements' index of the *Array* 9.

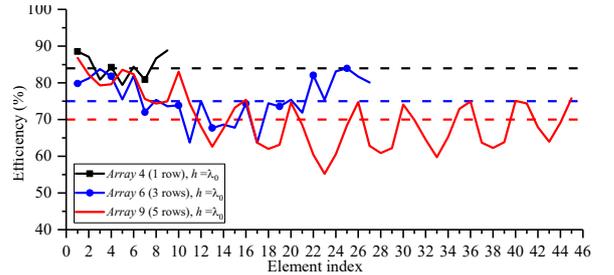

Fig. 27. Element efficiency of *Array* 4, *Array* 6 and *Array* 9 at 825 MHz.

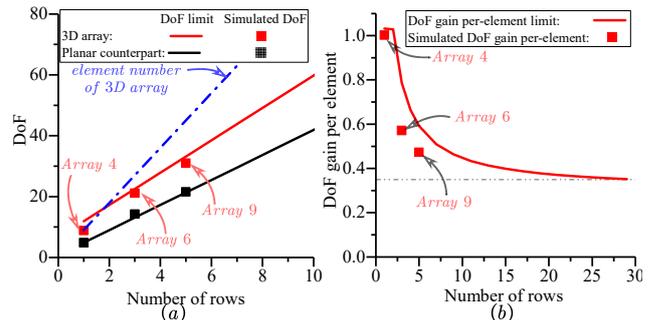

Fig. 28. (a) DoF and (b) Dof gain per-element of *Array* 4, *Array* 6 and *Array* 9 at 825 MHz.

To investigate the effect of array scaling on the DoF, Fig. 28(a) compares the DoF of *Array* 4, *Array* 6, and *Array* 9 (with 1, 3 and 5 rows, respectively) for an upper element profile of $h=\lambda_0$. Additionally, the theoretical half-space DoF limits for both 3D and planar arrays with varying numbers of rows are calculated using [21],

$$DoF\ limit = \frac{1}{4}\frac{k^2 A}{2\pi} \qquad (9)$$

where $k$ is the wavenumber and $A$ is the surface area of the smallest convex polyhedron enclosing the array, as illustrated in Fig. 29(a). When the upper element profile and number of rows are fixed, $A$ increases linearly with the number of columns. As shown in Fig. 28(a), the DoF of *Array* 4, *Array* 6, and *Array* 9 closely approach their respective theoretical limits. Specifically, *Array* 6 achieves a half-space DoF of 21.2, compared to the theoretical limit of 22.5, while *Array* 9 achieves 31 versus a limit of 33.2. These results demonstrate that using miniaturized elements enables the array to effectively approach the theoretical upper bound of DoF.

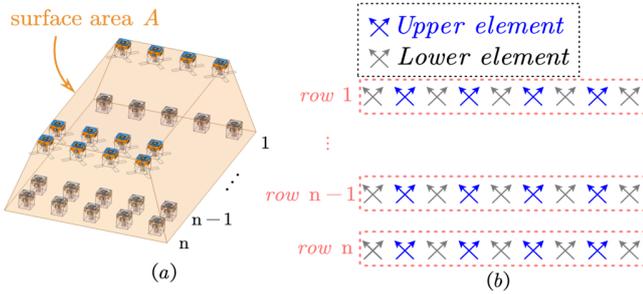

Fig. 29. (a) Surface area for DoF limit calculation, (b) Array typology with different number of rows.

However, as the array scale increases, the DoF grows at a slower rate than the number of elements—i.e., the slope of the theoretical DoF curve is shallower than that of the element count—indicating a reduction in DoF gain per-element. This DoF gain per-element is quantified using

$$DoF\ gain\ per\text{-}element = \frac{DoF_{3D} - DoF_{planar}}{N_{3D} - N_{planar}} \qquad (10)$$

, where $DoF_{3D}$ and $DoF_{planar}$ denote the DoF of the 3D array and its planar counterpart, respectively, while $N_{3D}$ and $N_{planar}$ represent their respective numbers of elements. Equation (10) characterizes the incremental DoF gain achieved by adding each element when transitioning from a planar to a 3D array. For instance, *Array* 4 achieves a DoF gain of 3.92 over its planar counterpart by adding 4 additional elements, yielding a DoF gain per-element of 0.98. Fig. 28(b) shows the variation of DoF gain per-element as the array scale increases. As observed, the DoF gain per-element for *Array* 6 and *Array* 9 are 0.57 and 0.47, respectively. Furthermore, by comparing the theoretical DoF limits of 3D and planar arrays, an upper bound on the DoF gain per-element can be derived. Although the actual DoF gains per-element for *Array* 4, *Array* 6, and *Array* 9 fall below this theoretical limit, they follow a consistent trend, indicating that the theoretical upper bound can serve as a reliable predictor for large-scale array configurations. As the array scale continues to increases, the DoF gain per-element asymptotically approaches a limiting value of approximately 0.35.

## V. MIMO Performance

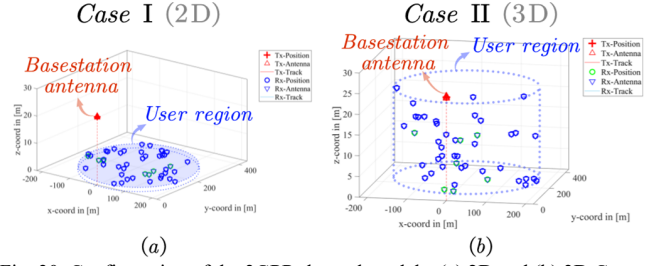

Fig. 30. Configuration of the 3GPP channel models: (a) 2D and (b) 3D Cases.

To further evaluate the MIMO performance advantages of 3D arrays in realistic communication environments, the QuaDRiGa software was employed to model both 2D and 3D urban macro (UMa) scenarios based on the 3GPP specifications. In this setup, base station antennas are mounted at a height of 25 meters and operate in transmit mode. Users are randomly distributed within a circular area with a radius of 200 meters. Two user distribution scenarios are considered, as illustrated in Fig. 30. In *Case* I, all users are located at a uniform height of 1.5 meters. In *Case* II, 80% of the users are distributed at varying heights, while the remaining 20% are at 1.5 meters. To generate the channel matrix, the active radiation patterns of *Array* 9 (with $h=\lambda_0$) and its planar counterpart are used as the transmit arrays. User terminals are equipped with omnidirectional, single-polarized antennas. MMSE precoding is assumed, and the capacity of each user is individually computed using

$$C^{(\chi)}(\mathbf{r}_u) = \log_2\left(1 + SINR^{(\chi)}(\mathbf{r}_u)\right) \qquad (11)$$

where

$$SINR^{(\chi)}(\mathbf{r}_u) = \frac{\left|\sum_{m=1}^{M}\sum_{n=1}^{N} w_{m,n}^{(u_i,\Psi)} \mathbf{G}^{(\Psi,\chi)}(\omega;\mathbf{r}_u,\mathbf{r}_{m,n})\right|^2}{\left|\sum_{u=1, u\neq u_i}^{U}\sum_{m=1}^{M}\sum_{n=1}^{N} w_{m,n}^{(u,\Psi)} \mathbf{G}^{(\Psi,\chi)}(\omega;\mathbf{r}_u,\mathbf{r}_{m,n})\right|^2 + \frac{U\sigma^2}{\Psi}} \qquad (12)$$

$\mathbf{G}^{(\Psi,\chi)}(\omega;\mathbf{r}_u,\mathbf{r}_{m,n})$ is the channel matrix with a dimension of $MN$-by-$U$ between the user locating at $\mathbf{r}_u$ and the antenna locating at $\mathbf{r}_{m,n}$. Notably, the channel matrix is normalized by adjusting its column vector moduli $|\mathbf{g}^{(\Psi,\chi)}(\omega;\mathbf{r}_u)|$ for each specific user $u$ individually [22], to match the realized gain of the antenna array in that user's direction $\mathbf{r}_u$. Besides, $w_{m,n}^{(u_i,\Psi)}$ is the beamforming coefficients, $\chi\in\{V,H\}$ stands for the polarizations, $\Psi$ is the input power of the transmit array, $U$ is the number of users while $\sigma^2$ is the noise power. The overall capacity of the MIMO system is then obtained by sum up the individual user's capacity.

$$C = \sum_{u=1}^{U}\sum_{\chi=V,H} C^{(\chi)}(\mathbf{r}_u) \qquad (13)$$

The resulting capacity under varying $\gamma = \Psi/\sigma^2$ for *Case* I is shown in Fig. 31. As observed, when $\gamma$=20 dB, the 3D array provides a 14% capacity gain over the planar array. Even under the extreme condition of $\gamma$=0 dB, the 3D array maintains a 6% improvement. The capacity performance in *Case* II is illustrated in Fig. 32. For the scenario with users distributed in a 3D region, the 3D array achieves a 16% capacity gain over the planar array at $\gamma$=20 dB, and a 23% gain at $\gamma$=30 dB. These results demonstrate that the miniaturized element 3D array offers substantial MIMO performance advantages in realistic propagation conditions.

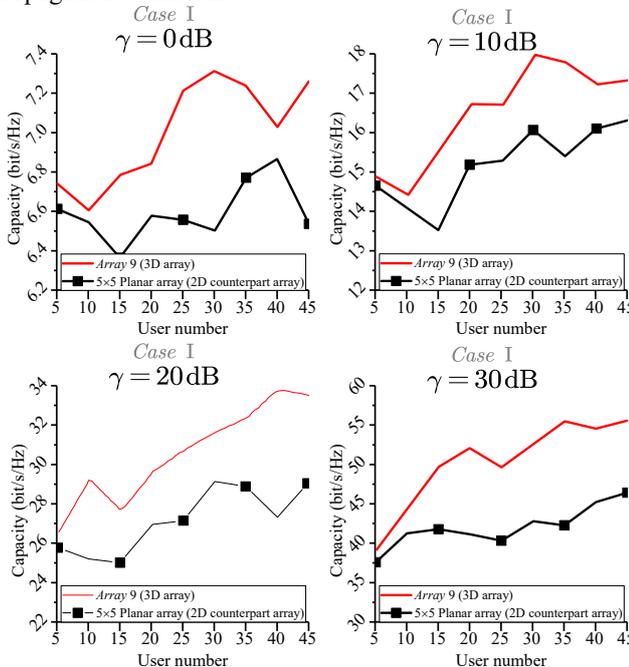

Fig. 31. Communication capacity of *Case* I (users distribute in 2D plane) related to user number under varies SNR.

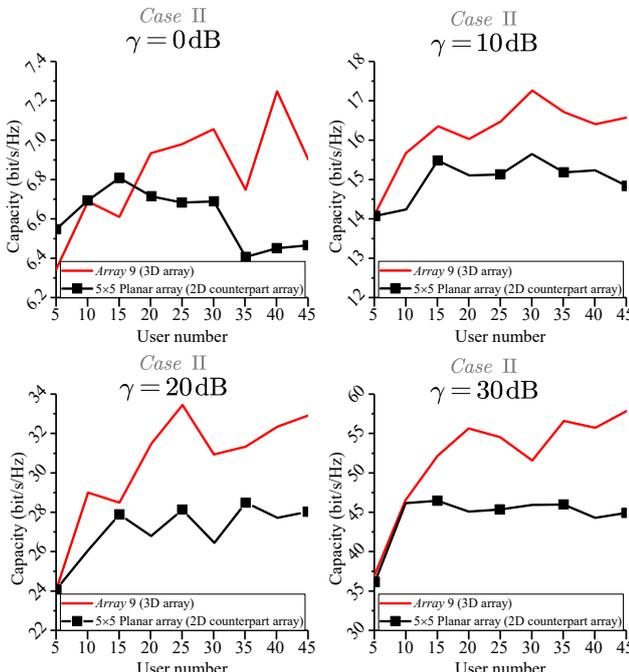

Fig. 32. Communication capacity of the *Case* II (users distribute in 3D region) related to user number under varies SNR.

Notably, the planar array's capacity has already reached the theoretical upper bound imposed by its physical aperture size, making further enhancement through techniques such as element decoupling or densification impractical. In contrast, the 3D array architecture benefits from increased DoF and higher efficiency, allowing for a capacity that exceeds the fundamental aperture-limited bounds. Furthermore, the proposed 3D arrays meet essential performance criteria for practical and industrial deployment, presenting a promising solution for future MIMO base station implementations.

## VI. Conclusion

In this work, a 3D antenna array based on miniaturized elements is proposed to address the challenges of DoF and efficiency degradation commonly observed in 3D arrays employing conventional antenna elements. The proposed array achieves a high DoF that exceeds the physical limit imposed by a planar aperture, while maintaining high array efficiency across a wide bandwidth. Furthermore, the effects of upper element profile distribution and the scalability of the proposed architecture for large-scale arrays are systematically analyzed. Finally, the communication capacity of the large-scale 3D array is evaluated using the 3GPP channel model. The results demonstrate that the miniaturized-element-based 3D array offers significant capacity improvements over its planar counterpart in MIMO scenarios.

## References

[1] R. He, B. Ai, G. L. Stuber, G. Wang, and Z. Zhong, "Geometrical-Based Modeling for Millimeter-Wave MIMO Mobile-to-Mobile Channels," *IEEE Trans. Veh. Technol.*, vol. 67, no. 4, pp. 2848–2863, Apr. 2018, doi: 10.1109/TVT.2017.2774808.
[2] D. Tse and P. Viswanath, *Fundamentals of Wireless Communication*, 1st ed. Cambridge University Press, 2005. doi: 10.1017/CBO9780511807213.
[3] P. Hannan, "The element-gain paradox for a phased-array antenna," *IEEE Trans. Antennas Propag.*, vol. 12, no. 4, pp. 423–433, Jul. 1964, doi: 10.1109/TAP.1964.1138237.
[4] A. Pizzo, T. L. Marzetta, and L. Sanguinetti, "Degrees of Freedom of Holographic MIMO Channels," in *2020 IEEE 21st International Workshop on Signal Processing Advances in Wireless Communications (SPAWC)*, Atlanta, GA, USA: IEEE, May 2020, pp. 1–5. doi: 10.1109/SPAWC48557.2020.9154219.
[5] G. Yang and S. Zhang, "Dual-Polarized Wide-Angle Scanning Phased Array Antenna for 5G Communication Systems," *IEEE Trans. Antennas Propag.*, vol. 70, no. 9, pp. 7427–7438, Sep. 2022, doi: 10.1109/TAP.2022.3141188.
[6] J. Jiang and Q.-X. Chu, "Broadband Decoupling for Antenna Arrays Using Multiple Decoupling Nulls," *IEEE Trans. Antennas Propag.*, vol. 71, no. 11, pp. 8616–8627, Nov. 2023, doi: 10.1109/TAP.2023.3312821.
[7] A. Zhang, K. Wei, and Z. Zhang, "Multiband and Wideband Self-Multipath Decoupled Antenna Pairs," *IEEE Trans. Antennas Propag.*, vol. 71, no. 7, pp. 5605–5615, Jul. 2023, doi: 10.1109/TAP.2023.3267177.
[8] Z. Zhou, Y. Ge, J. Yuan, Z. Xu, and Z. D. Chen, "Wideband MIMO Antennas With Enhanced Isolation Using Coupled CPW Transmission Lines," *IEEE Trans. Antennas Propag.*, vol. 71, no. 2, pp. 1414–1423, Feb. 2023, doi: 10.1109/TAP.2022.3232693.
[9] S. Jaeckel, L. Raschkowski, K. Borner, and L. Thiele, "QuaDRiGa: A 3-D Multi-Cell Channel Model With Time Evolution for Enabling Virtual Field Trials," *IEEE Trans. Antennas Propag.*, vol. 62, no. 6, pp. 3242–3256, Jun. 2014, doi: 10.1109/TAP.2014.2310220.
[10] M. Li, X. Chen, A. Zhang, A. A. Kishk, and W. Fan, "Reducing Correlation in Compact Arrays by Adjusting Near-Field Phase Distribution for MIMO Applications," *IEEE Trans. Veh. Technol.*, vol. 70, no. 8, pp. 7885–7896, Aug. 2021, doi: 10.1109/TVT.2021.3094314.
[11] X. Chen *et al.*, "Simultaneous Decoupling and Decorrelation Scheme of MIMO Arrays," *IEEE Trans. Veh. Technol.*, vol. 71, no. 2, pp. 2164–2169, Feb. 2022, doi: 10.1109/TVT.2021.3134180.
[12] Y. Wang *et al.*, "Improvement of Diversity and Capacity of MIMO System Using Scatterer Array," *IEEE Trans. Antennas Propag.*, vol. 70, no. 1, pp. 789–794, Jan. 2022, doi: 10.1109/TAP.2021.3098568.

[13] S. S. A. Yuan *et al.*, "Breaking the Degrees-of-Freedom Limit of Holographic MIMO Communications: A 3-D Antenna Array Topology," *IEEE Trans. Veh. Technol.*, vol. 73, no. 8, pp. 11276–11288, Aug. 2024, doi: 10.1109/TVT.2024.3372704.

[14] M. D. Migliore, "Horse (Electromagnetics) is More Important Than Horseman (Information) for Wireless Transmission," *IEEE Trans. Antennas Propag.*, vol. 67, no. 4, pp. 2046–2055, Apr. 2019, doi: 10.1109/TAP.2018.2889158.

[15] X. Chen, P.-S. Kildal, J. Carlsson, and J. Yang, "MRC Diversity and MIMO Capacity Evaluations of Multi-Port Antennas Using Reverberation Chamber and Anechoic Chamber," *IEEE Trans. Antennas Propag.*, vol. 61, no. 2, pp. 917–926, Feb. 2013, doi: 10.1109/TAP.2012.2223442.

[16] P.-S. Kildal and K. Rosengren, "Correlation and capacity of MIMO systems and mutual coupling, radiation efficiency, and diversity gain of their antennas: simulations and measurements in a reverberation chamber," *IEEE Commun. Mag.*, vol. 42, no. 12, pp. 104–112, Dec. 2004, doi: 10.1109/MCOM.2004.1367562.

[17] S. Biswas, C. Masouros, and T. Ratnarajah, "Performance Analysis of Large Multiuser MIMO Systems With Space-Constrained 2-D Antenna Arrays," *IEEE Trans. Wirel. Commun.*, vol. 15, no. 5, pp. 3492–3505, May 2016, doi: 10.1109/TWC.2016.2522419.

[18] J. B. Andersen and K. I. Pedersen, "Angle-of-arrival statistics for low resolution antennas," *IEEE Trans. Antennas Propag.*, vol. 50, no. 3, pp. 391–395, Mar. 2002, doi: 10.1109/8.999632.

[19] M. T. Ivrlac and J. A. Nossek, "Diversity and Correlation in Rayleigh Fading MIMO Channels," in *2005 IEEE 61st Vehicular Technology Conference*, Stockholm, Sweden: IEEE, 2005, pp. 151–155. doi: 10.1109/VETECS.2005.1543268.

[20] W. Yang, Y. Li, Q. Xue, S. Liao, and W. Che, "Miniaturized Broadband Dual-Polarized Dipole Antenna Based on Multiple Resonances and its Array for Base-Station Applications," *IEEE Trans. Antennas Propag.*, vol. 70, no. 11, pp. 11188–11193, Nov. 2022, doi: 10.1109/TAP.2022.3209273.

[21] M. Gustafsson and J. Lundgren, "Degrees of Freedom and Characteristic Modes: Estimates for radiating and arbitrarily shaped objects," *IEEE Antennas Propag. Mag.*, vol. 66, no. 6, pp. 18–28, Dec. 2024, doi: 10.1109/MAP.2024.3389451.

[22] S. S. A. Yuan, L. Wei, X. Chen, C. Huang, and W. E. I. Sha, "Electromagnetic Normalization of Channel Matrix for Holographic MIMO Communications," *IEEE Trans. Wirel. Commun.*, pp. 1–1, 2025, doi: 10.1109/TWC.2025.3543585.